\documentclass[useAMS]{mn2e}
\usepackage{graphicx}
\begin{document}
\def\simlt{\mathrel{\rlap{\lower 3pt\hbox{$\sim$}}
        \raise 2.0pt\hbox{$<$}}}
\def\simgt{\mathrel{\rlap{\lower 3pt\hbox{$\sim$}}
        \raise 2.0pt\hbox{$>$}}}
\def\bj{b_{\rm\scriptscriptstyle J}}
\def\rt{r_{\rm\scriptscriptstyle T}}
\def\rp{r_{\rm\scriptscriptstyle P}}
\renewcommand{\labelenumi}{(\arabic{enumi})}
\title[On the evolution of clustering of 24$\mu$m-selected galaxies.]
{On the evolution of clustering of 24$\mu$m-selected galaxies.}
\author[Manuela Magliocchetti et al.]
{\parbox[t]\textwidth{M. Magliocchetti$^{1,2,3}$, M.Cirasuolo$^{4}$, 
R.J. McLure$^{4}$, J.S. Dunlop$^{4}$, O. Almaini$^{5}$, S. Foucaud$^{5}$,
G. De Zotti$^{6,3}$, C. Simpson$^{7}$, K. Sekiguchi$^{8}$}\\
\tt $^1$ INAF, Osservatorio Astronomico di Trieste, Via Tiepolo 11, 34100,
Trieste, Italy\\
\tt $^2$ ESO, Karl-Schwarzschild-Str.2, D-85748, Garching, Germany\\
\tt $^3$ SISSA, Via Beirut 4, 34014, Trieste, Italy\\
\tt $^4$ SUPA, Scottish Universities Physics Alliance, 
Institute for Astronomy, University of Edinburgh, \\\tt \hspace{0.4cm}Royal Observatory, 
Edinburgh, EH9, 3HJ, UK\\
\tt $^5$ School of Physics and Astronomy, University of Nottingham,
University Park, Nottingham NG7 2RD, UK\\ 
\tt $^6$ INAF, Osservatorio Astronomico di Padova, Vicolo dell'Osservatorio 5, 35122
Padova, Italy\\
\tt $^7$ Astrophysics Research Institute, Liverpool John Moores University, Twelve
Quays House, Egerton Wharf, Birkenhead CH41 1LD, UK\\
\tt $^8$ National Astronomical Observatory of Japan, Mitaka, Tokyo 181-8588, Japan\\}
\maketitle

\vspace{7cm}
\begin{abstract}
This paper investigates the clustering properties of a complete sample of 1041 
24$\mu$m-selected sources brighter than $F_{24\mu \rm m}=400\mu$Jy in the
overlapping region between the SWIRE and UKIDSS UDS surveys. With the help 
of photometric redshift determinations we have concentrated on the two interval ranges 
$z=[0.6-1.2]$ ({\it low-z} sample) and $z\ge 1.6$ ({\it high-z} sample) as it is in these regions 
were we expect the mid-IR population to be dominated by intense dust-enshrouded activity such 
as star formation and black hole accretion. Investigations of the angular correlation 
function produce an amplitude $A\sim 0.010$ for the high-z sample and $A\sim 0.0055$ for the 
low-z one. The corresponding correlation lengths are $r_0=15.9^{+2.9}_{-3.4}$~Mpc and 
$r_0=8.5^{+1.5}_{-1.8}$~Mpc, showing that the high-z population is more strongly clustered.
Comparisons with physical models for the formation and evolution of large-scale structure 
reveal that the high-z sources are exclusively associated with very massive ($M\simgt 10^{13} M_\odot$)
haloes, comparable to those which locally host groups-to-clusters of galaxies, and are very common 
within such (rare) structures. Conversely, lower-z galaxies are found to reside in smaller halos 
($M_{\rm min}\sim 10^{12} M_\odot$) and to be very rare in such systems.  On the other hand, mid-IR 
photometry shows that the low-z and high-z samples include similar objects and probe a similar 
mixture of AGN and star-forming galaxies.
While recent studies have determined a strong evolution of the 24$\mu$m luminosity function 
between $z\sim 2$ and $z\sim 0$, they cannot provide information on the 
physical nature of such an evolution. Our clustering results instead  
indicate that this is due to the presence of different populations of 
objects inhabiting different structures, as active systems at $z\simlt 1.5$ 
are found to be exclusively associated with low-mass galaxies, while very massive sources appear to have concluded 
their active phase before this epoch. 
Finally, we note that the small-scale clustering data seem to require steep ($\rho\propto r^{-3}$) 
profiles for the distribution of galaxies within their halos. This is suggestive of close encounters 
and/or mergers which could strongly favour both AGN and star-formation activity.
\end{abstract}



\section{Introduction}
Understanding the assembly history of massive spheroidal galaxies
is a key issue for galaxy formation models.  The ``naive"
expectation from the canonical hierarchical merging scenario, that
proved to be remarkably successful in explaining many aspects of
large-scale structure formation, is that massive galaxies
generally form late and over a long period of time as the result
of many mergers of smaller haloes. On the other hand, there is
quite extensive evidence that massive galaxies may form at high
redshifts and on short timescales (see, e.g. Cimatti et al. 2004;
Fontana et al. 2004; Glazebrook et al. 2004; 
Treu et al. 2005; Saracco et al. 2006; Bundy et al. 2006; McLure et al. 2006),
while the sites of active star formation shift to lower mass
systems at later epochs, a pattern referred to as "downsizing"
(Cowie et al. 1996; Heavens et al. 2004). In order to reconcile the observational evidence
that stellar populations in large
spheroidal galaxies are old and essentially coeval (Ellis et al.
1997; Holden et al. 2005) with the hierarchical
merging scenario, the possibility of mergers of evolved
sub-units (``dry mergers'') has been introduced (van Dokkum et al.
2005; Naab et al. 2006). This mechanism is
however strongly disfavoured by studies on the evolution of the
stellar mass function (Bundy et al. 2006).

Key information, complementary to optical/IR data, has come from
sub-millimeter surveys (Hughes et al. 1998; Eales et al. 2000;
Scott et al. 2002; Knudsen et al. 2006; Coppin et al. 2006) which have found 
a large population of
luminous sources at substantial redshifts (Chapman et al. 2005).
However, the interpretation of this class of objects is still
controversial (e.g. Granato et al. 2004; Kaviani et al. 2003; Baugh et al. 2005).
The heart of the problem are the masses of the objects: a large
fraction of present day massive galaxies already assembled at
$z\sim 2-3$ would be extremely challenging for the standard view
of a merging-driven  growth.   Measurements of clustering
amplitudes are a unique tool to estimate halo masses at high $z$,
but complete samples comprising at least several hundred of
sources are necessary.

Recently, Magliocchetti et al. (2007) have reported evidence for 
strong clustering for $\sim 800$ optically very faint ($R>25.5$), $F_{24\mu \rm m}\ge 0.35$~mJy 
sources obtained from the Spitzer first cosmological survey (First Look Survey -- FLS; 
Fadda et al. 2006). Both the clustering properties and the counts of
such sources are consistent with them being very massive
proto-spheroidal galaxies in the process of forming most of their
stars. Furthermore, by assuming a medium redshift $z\sim 2$, their comoving number  
density appears to be much higher than what expected from most semi-analytic models.

The Magliocchetti et al. (2007) work however suffers from the lack of information on the redshift 
distribution of the optically faint Spitzer-selected sources and has to rely on models 
based on both template spectral energy distributions and on theoretical investigations of the issue of 
galaxy formation and evolution in order to go from the observed projected clustering signal to the more 
meaningful results in real space.

The optical-to-mid-IR depth of the UKIDSS data allows us to overcome this problem. 
Photometric estimates for the overwhelming majority ($\sim 97$\%) of 24$\mu$m-selected sources 
are now available (Cirasuolo et al. 2007). Furthermore, despite the somehow 
poor statistics, the redshift information can also 
allow us for the first time to compare 
the clustering signal of similar sources at different epochs so to investigate possible 
differences and evolution in their large-scale properties.    
To this aim, we will concentrate on two samples, the first one which includes galaxies in the 
$z=[0.6-1.2]$ range, and a second one made by sources with $z\simgt 1.6$. Diagnostics based 
on mid-IR photometry indicate that these two samples are likely made by a very similar mixture of 
active star-forming galaxies and AGN.

The layout of the paper is as follows: In \S\,2 we describe the parent catalogue and the sample 
selection.  In \S 3.1 we derive the two point
angular correlation function, while in \S 3.2 we present the results for the spatial clustering 
properties of the sources in exam.
\S 4 discusses the implications of the clustering results on the source properties, and in particular for 
what concerns their halo masses and number density.
Our main conclusions are summarized in \S\,5.

Throughout this work we adopt a flat cosmology with $\Omega_m=0.3$ and 
$\Omega_\Lambda=0.7$, a present-day value of the Hubble parameter
in units of $100$ km/s/Mpc $h=0.7$, and rms density fluctuations
within a sphere of $8 h^{-1}$ Mpc radius $\sigma_8=0.8$ (Spergel
et al. 2003).

\section {Sample Selection}

For this work we used the {\it Spitzer} wide-area infrared extragalactic 
(SWIRE) survey (Lonsdale et al. 2003; 2004) to select sources with fluxes at 24
$\mu$m brighter than 400 $\mu$Jy ($5\sigma$ completeness)
in the XMM-LSS field (Surace et al. 2005). In order to obtain multi-wavelength
information and accurate redshift estimates for these sources we limited our analysis
to the region of the SWIRE survey which overlaps with the 0.7 square degrees
covered by the UKIDSS Ultra Deep Survey (UDS - Lawrence et al. 2006) 
and Subaru imaging (Sekiguchi et al. 2005; Furusawa et al. in
preparation).  The 5 overlapping Subaru 
Suprime-Cam pointings provide broad band photometry in the $BVRi'z'$ filters to typical 
$5 \sigma$ depths of $B=27.5$, $V=26.7$, $R=27.0$, $i'=26.8$ and $z'=25.9$ 
(within a $2$-arcsec diameter aperture). The UDS is the deepest of the five
surveys that constitute the UKIDSS survey (Lawrence et al. 2006) and for this
work we used $J$ and $K$-band imaging from the first data release with a
$5 \sigma$ depth of $J_{AB} = 23.5$  and $K_{AB} = 23.5$ (Warren et
al. 2007). Finally, the SWIRE survey also provides Spitzer-IRAC measurements 
at 3.6, 4.5, 5.8 and 8$\mu$m (Surace et al. 2005). 

\begin{figure}
\includegraphics[width=8.0cm]{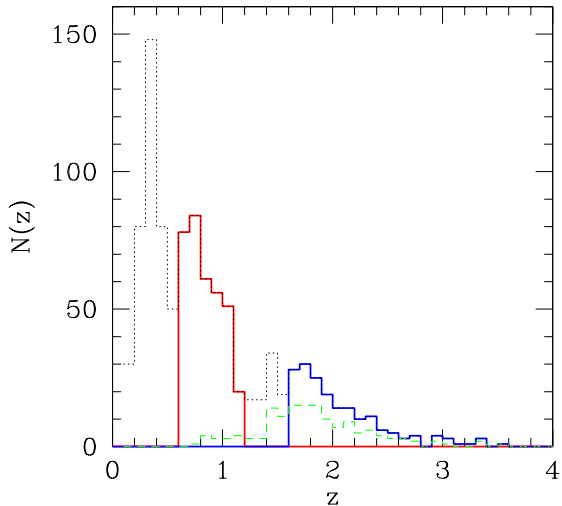} 
\caption{Photometric redshift distribution of sources in the UDS-SWIRE, 
$F_{24\mu \rm m}\ge 400 \mu$Jy 
sample (thin dotted line). The two thick solid lines highlight the redshift 
intervals covered by the high-z (blue) and low-z (red) samples. 
For comparison, the (green) dashed line indicates the redshift distribution of 
all $F_{24\mu \rm m}\ge 400 \mu$Jy sources fainter than $R=24.5$. No redshift 
cut was performed on this last sample.
\label{figure:nz}}
\end{figure}

Within the overlapping region between SWIRE and UDS we isolated
1184 sources with $F_{24\mu \rm m}\ge 400\mu$Jy, out of which 1041 have a reliable 
optical and/or near-infrared counterpart. We take these objects as our working sample,
UDS-SWIRE hereafter.
As extensively described in Cirasuolo et al. (2007), photometric redshifts for these sources 
have been computed by fitting the 
observed photometry (9 broad bands from 0.4 to 4.5$\mu m$) 
with both synthetic (Bruzual \& Charlot 2003) and empirical (Coleman, Wu \& Weedman 1980; 
Kinney et al. 1996; Mignoli et al. 2005) galaxy templates, by using the public package 
{\sc HYPERZ} (Bolzonella, Miralles \& Pell\'{o} 2000).
Comparisons with spectroscopic redshifts available 
in the field show the good accuracy of the photometric redshifts with a  
$\Delta z /(1+z) \simeq 0.05$ over the redshift range $0<z<6$ (Cirasuolo et al. 2007).\\ 
The redshift distribution $N(z)$ for our 24$\mu$m sources is 
shown by the dotted line in  Figure~{\ref{figure:nz}}. 
As it is possible to appreciate from the Figure, 
the objects included in the UDS, $F_{24\mu \rm m}\ge 400\mu$Jy 
sample are found up to 
redshifts $z\sim 3.5$. \\

\noindent
In order to investigate possible evolutionary features of the 
large-scale structure as traced by these objects, we have concentrated 
on two sub-samples: a first one 
which includes both sources with redshifts above 1.6 and those without 
an optical or K-band identification (hereafter called the 
{\it high-z} sample) and a second one with objects within $z=0.6$ and $z=1.2$ 
(hereafter called the {\it low-z} sample). 
There are a number of reasons for the above choice. 
$z\sim 1.6$ is in fact the redshift at which the strong spectral PAH feature centred at 7.7$\mu$m -- 
indicative of a very intense star-forming activity -- enters 
the 24$\mu$m band. The high-z sample is therefore expected to include 
a relevant fraction of galaxies 
undergoing an intense phase of dust-enshrouded stellar formation. 
Furthermore, recent studies have proved that a strong 24$\mu$m emission 
combined with a very faint or no optical detection most likely originates from 
obscured star-forming galaxies at redshifts between about 1.6 and 2.7 
(see e.g. Yan et al. 2005; 2007; Houck et al. 2005). On the other hand, 
results both based on IRS spectroscopy and on diagnostics of the 
ratio between 8$\mu$m and 24$\mu$m fluxes, have inferred a fraction of $z\sim 2$ obscured 
AGN brighter than our chosen 24$\mu$m flux limit of about 30\%, 
figure which rapidly increases to $\sim 100$\% at the highest 24$\mu$m fluxes 
(see e.g. Brand 2006; Magliocchetti et al. 2007). The counts presented in the right-hand panel of Figure~2 
seem to confirm this expectation ($\sim 65$\% candidate starburst galaxies on the basis of 
their $F_{8\mu \rm m}/F_{24\mu \rm m}< 0.1$ ratios, with 
the remaining $\sim 35$\% being made of candidate AGN-dominated objects).

\begin{figure*}
\includegraphics[width=8.0cm]{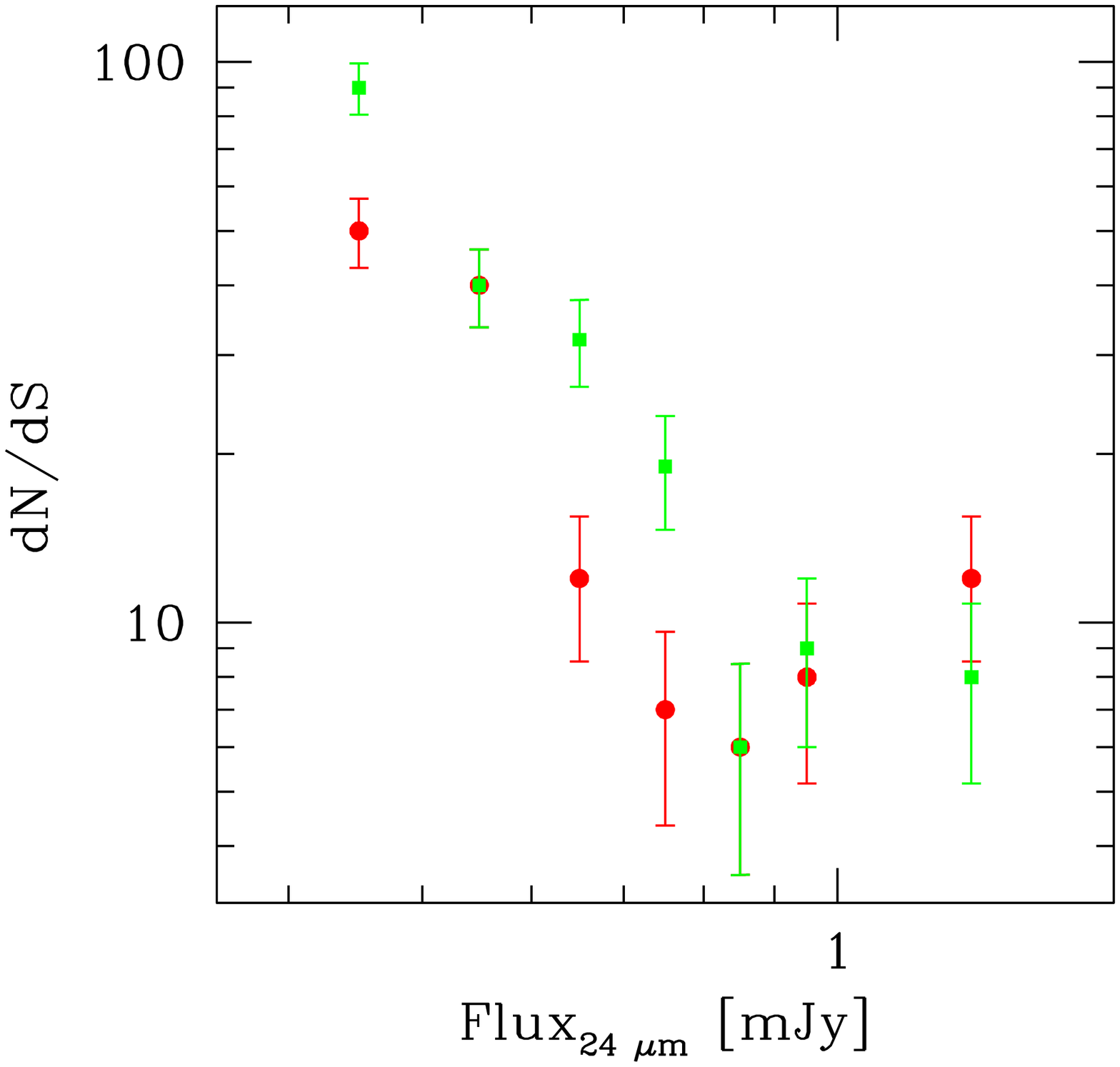} 
\includegraphics[width=8.0cm]{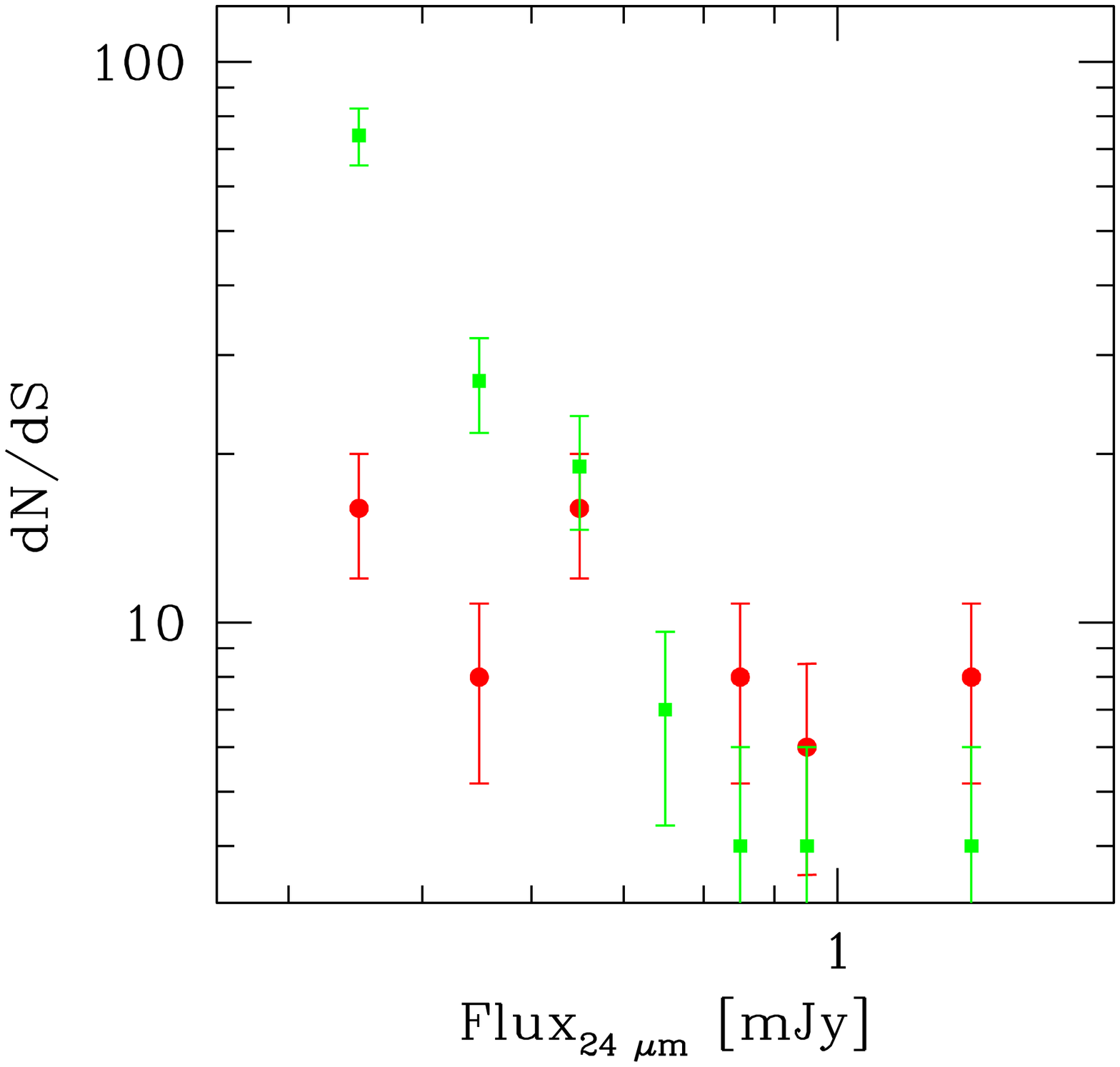}
\label{figure:N_S}
\caption{Differential number counts for candidate star-forming galaxies 
($F_{8\mu\rm m}/F_{24\mu\rm m}< 0.1$; green squares) and candidate AGN-dominated sources 
($F_{8\mu\rm m}/F_{24\mu\rm m}\ge 0.1$; red circles) 
in the low-z (left-hand panel) and high-z (right-hand panel) samples. }
\end{figure*}

\begin{figure*}
\includegraphics[width=8.0cm]{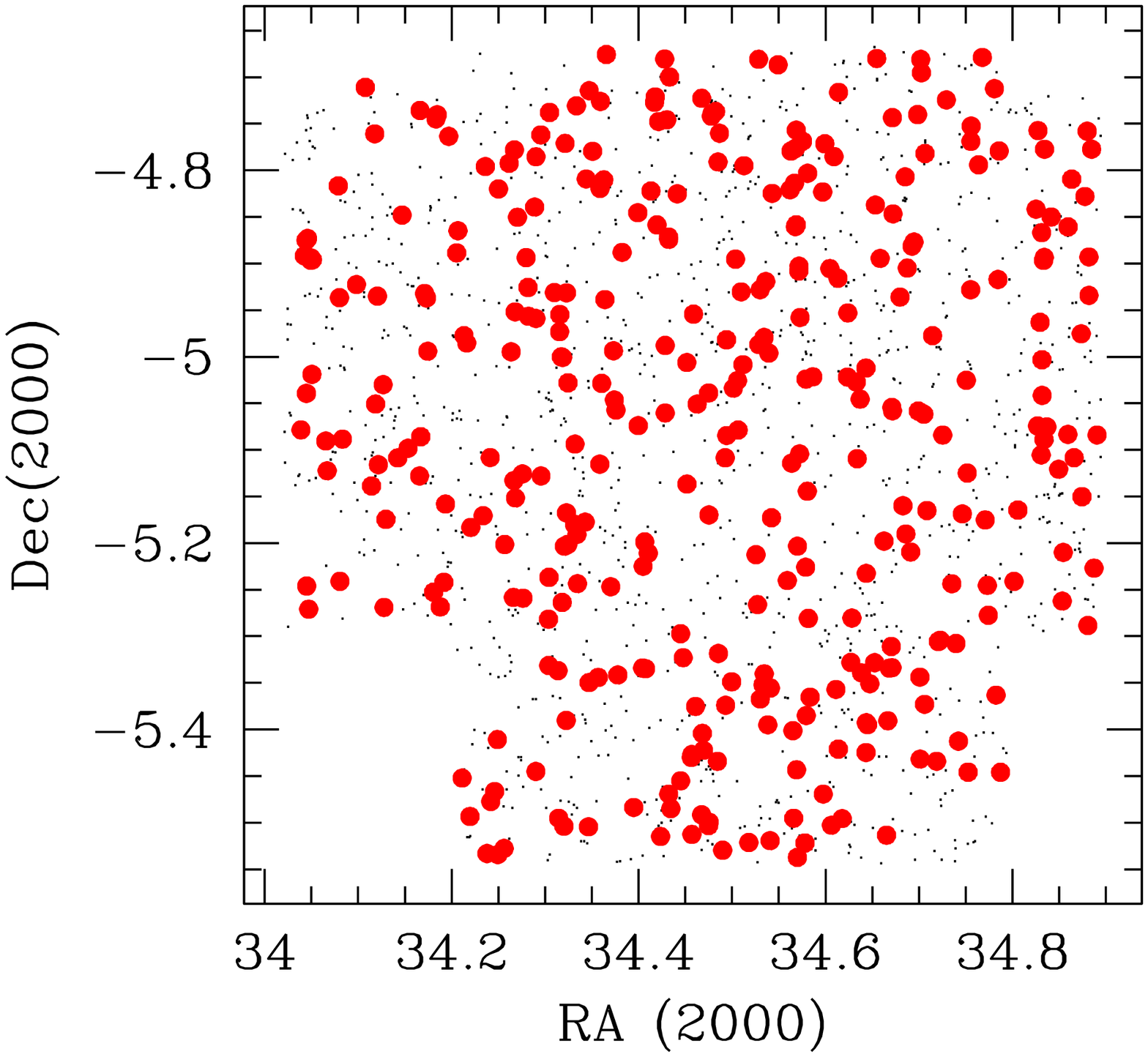} 
\includegraphics[width=8.0cm]{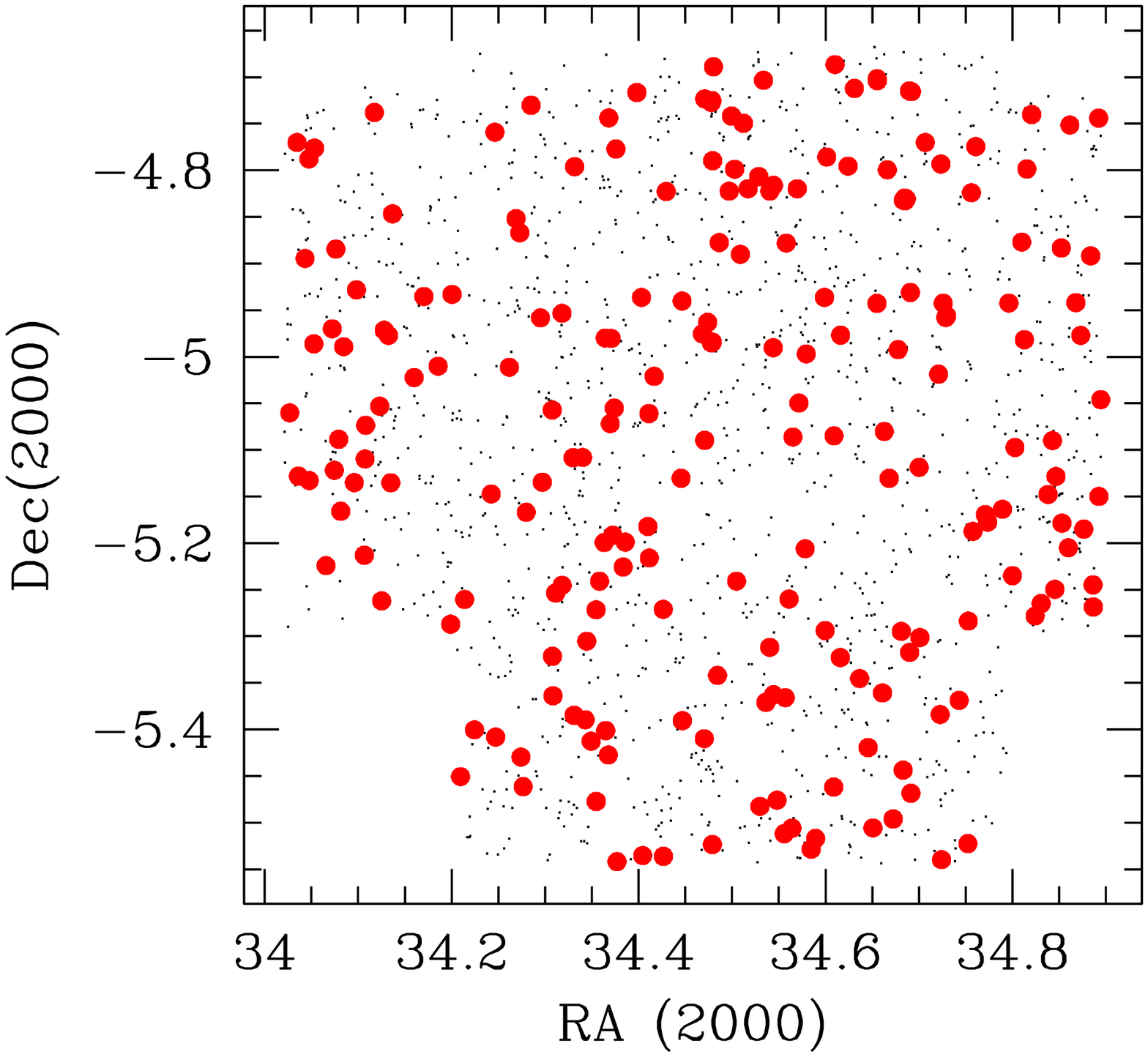}
\label{figure:radec}
\caption{Sky distribution of UDS-SWIRE, $F_{24\mu \rm m}\ge 400 \mu$Jy sources 
(small dots). The filled (red) circles in the left-hand panel illustrate the 
350 objects included in the redshift interval $z=[0.6-1.2]$, while those 
in the right-hand panel represent the 210 sources with either $z>1.6$ or no 
optical or near-IR identification.}
\end{figure*}

The low-z sample is also expected to probe a substantial fraction of obscured 
star-forming galaxies and AGN, albeit with lower intrinsic luminosities since we are 
dealing with a flux-limited survey. 
The $24\mu\rm m$ counts presented in the left-hand 
panel of Figure~2 are remarkably similar to those obtained for the high-z sample 
($\sim 60$\% and $\sim 40$\% respectively for candidate starburst galaxies and 
candidate AGN-dominated objects; we note that the lower limit of $z\sim 0.6$ 
adopted for the low-z sample corresponds to the lowest redshift at which a meaningful distinction 
between AGN-powered and SF-dominated sources as based on the ratio between $F_{8\mu\rm m}$ and 
$F_{24\mu\rm m}$ fluxes can be made given the spectral properties of intense star-forming galaxies in 
the mid-IR). Furthermore, from recent multi-photometry studies 
which extend from the optical to the X-ray band, it seems that $z\sim 1$ 
corresponds to the 'bulk' of obscured AGN activity 
(see e.g. Pozzi et al. 2007). A comparative analysis of the two samples 
would then allow to investigate any redshift evolution in the large-scale 
properties of these sources and, if present, find an evolutionary 
connection between objects in the low-z and high-z samples. 

The {\it high-z} sample includes 210 sources, 28 of which do not have an optical 
counterpart. The redshift distribution of the objects with 
assigned photo-z closely resembles that of $R\ge 24.5$,
24$\mu$m-selected galaxies (cfr. Figure~\ref{figure:nz}), showing once again 
that optically-faint sources selected at 24$\mu$m most likely reside at 
redshifts $z > 1.6$. The projected distribution onto the sky 
of the high-z sources is shown by the filled (red) 
circles on the right-hand side of Figure~3. Small dots 
indicate the positions of all $F_{24\mu \rm m}\ge 400 \mu$Jy UDS-SWIRE objects.\\
By making use of the redshift distribution shown in Figure~\ref{figure:nz}, 
and assuming that the $N(z)$ of optically obscured sources follows that of galaxies 
with estimated redshift, 
we find that the average redshift for the above sample is $<z>=2.02$, 
its median is $z_{\rm med}=1.93$, and the 
mean comoving number density of such sources is $\bar{n}_G=(2.5 \pm 0.3) 
\cdot 10^{-5}$ Mpc$^{-3}$, number which decreases to 
$\bar{n}_G=(2.2 \pm 0.3)\cdot 10^{-5}$ Mpc$^{-3}$ if one only includes objects with 
redshifts. The errors associated to the above quantities 
represent the 2$\sigma$ confidence level as derived from Poisson statistics.
Estimates of the number density at $z\sim 2$ might be very tricky due to joint effect of 
the cosmological evolution of the sources under examination and to the large variance of the 
observed 24$\mu$m SED in that redshift range. This is why, for a safety check, we have also repeated 
our calculations of 
$\bar{n}_G$ by simply considering a redshift box extending from $z=1.6$ to $z=2.5$. The result is 
virtually identical to that quoted above both by including and excluding galaxies with no redshift 
information and also very similar to that obtained by extrapolating the $z\sim 2$ 
Caputi et al. (2007; cfr. their Figure 8) Luminosity Function for 24$\mu$m-selected galaxies brighter than 
$\nu L_{\nu}\simeq 10^{11.9} L_{\odot}$, 
where the limiting luminosity has been calculated for  $F_{24\mu\rm m}=0.4$~mJy 
by following an Arp220-like Spectral Energy Distribution, indicative of galaxies undergoing intense 
star-formation which we expect (cfr. earlier in this section and also Figure~2) to dominate our sample. 
 
The {\it low-z} sample instead contains 350 sources. Their sky distribution is 
represented by the filled (red) circles on the left-hand side of 
Figure~3. 
The average redshift of the sample is found to be $<z>=0.79$, its median $z_{\rm med}=0.84$, 
and the comoving number density of these sources is $\bar{n}_G=(9.9 \pm 1.0) 
\cdot 10^{-5}$ Mpc$^{-3}$.  Even in this case, the number density is in excellent agreement with 
the findings of Caputi et al. (2007) for their $z\sim 1$ sources with $\nu L_{\nu}\simgt 10^{11.3} 
L_{\odot}$, where $\nu L_{\nu}$ was calculated as indicated before. The most relevant properties for 
the high-z and low-z samples are summarized in Table~2.

\section{Clustering properties}
\subsection{The Angular Correlation Function}

The angular two-point correlation function
$w(\theta)$ gives the excess probability, with respect to a
random Poisson distribution, of finding two sources in the solid
angles $\delta\Omega_1$ $\delta\Omega_2$ separated by an angle
$\theta$. In practice, $w(\theta)$ is obtained by comparing the
actual source distribution with a catalogue of randomly
distributed objects subject to the same mask constraints as the
real data. We chose to use the estimator (Hamilton 1993)
\begin{eqnarray}
w(\theta) = 4\times \frac{DD\cdot RR}{(DR)^2} -1, 
\label{eq:xiest}
\end{eqnarray}
where $DD$, $RR$ and $DR$ are the number of data-data, random-random 
and data-random pairs separated by a distance $\theta$. 

\begin{figure*}
\includegraphics[width=8.0cm]{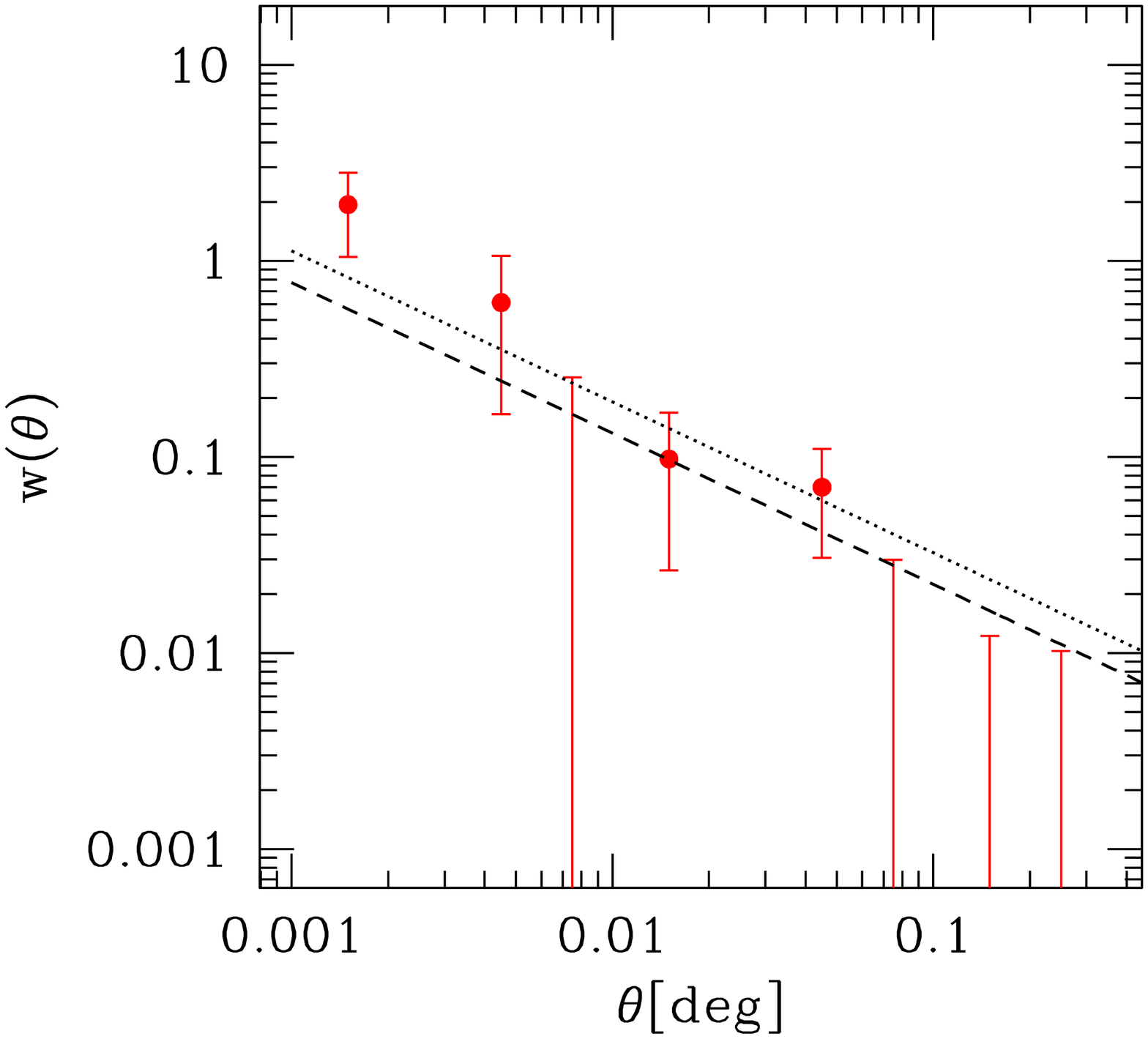} 
\includegraphics[width=8.0cm]{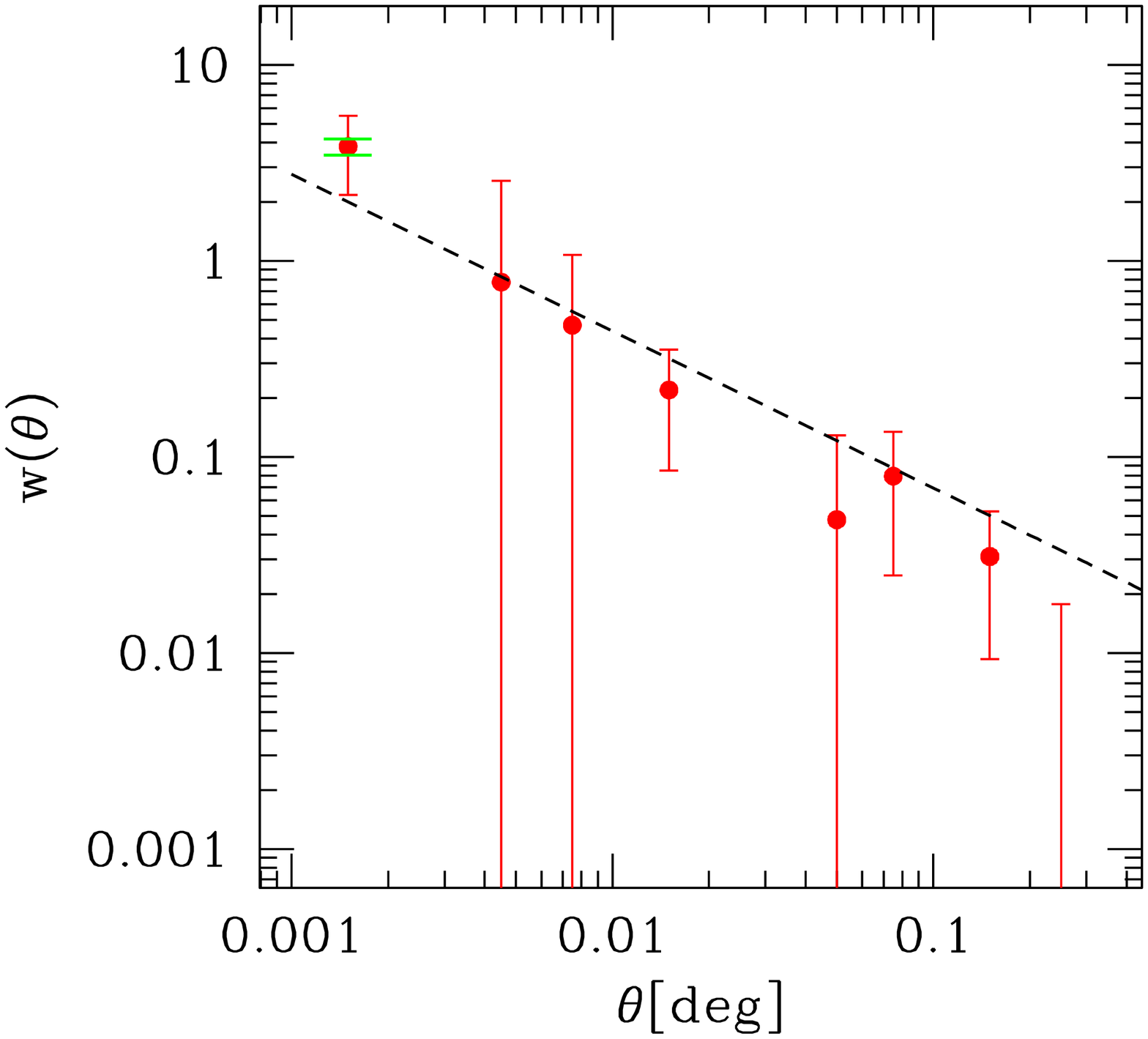}
\label{figure:w}
\caption{Angular correlation function $w(\theta)$ for the low-z 
(left-hand panel), and high-z (right-hand panel) samples. The dashed lines show the 
best power-law fits to the data, while the dotted line in the left-hand panel indicates 
the variation in the best fit obtained by considering the $w(\theta)+1\sigma$ values at those 
scales where $w(\theta)$ is negative as mere upper limits. 
The two horizontal (green) dashes in the plot for the high-z sample represent the small-scale variation 
of $w(\theta)$ derived by either including or excluding the pair 
of sources whose distance from each other is $<5.4''$ (see text for details).}
\end{figure*}

The two considered UDS-SWIRE samples have an estimated 
5$\sigma$ flux completeness $F_{24\mu \rm m}\ge 400 \mu$Jy. Furthermore, since the 
whole UDS area was uniformly observed by Spitzer/MIPS, 
we only had to mask out those (few) portions of the sky contaminated by the presence of 
bright stars. 
Random catalogues covering the whole surveyed area minus the bright star regions 
and with twenty times 
as many objects as the real data sets were then generated for both 
the high-z and low-z samples.
$w(\theta)$ in eq. (\ref{eq:xiest}) was then estimated in the angular range 
$10^{-3}\simlt \theta \simlt 0.3$ degrees. The upper limit is determined by 
the geometry of UDS, and corresponds to about half of the maximum scale 
probed by the survey. 

Figure 4 presents our results for the two low-z (left-hand panel) 
and high-z (right-hand panel) samples, while Table~1 reports the $w(\theta$) values  
as a function of angular scale in both cases. 
The error bars show Poisson
estimates for the points. Since the distribution is clustered, these estimates
only provide a lower limit to the uncertainties. However, it can be shown  
that over the considered range of angular scales this estimate is close to 
those obtained from bootstrap resampling (e.g. Willumsen, Freudling 
\& Da Costa, 1997).

Particular attention was devoted to the issue of close pairs. 
In fact, the points on the top left-hand corners of both $w(\theta)$ 
estimates correspond to angular scales close to the 5.4$^{\prime\prime}$ 
resolution of the 24$\mu$m Spitzer channel  and therefore could be affected
by source confusion. However, since optical and near-infrared images for these 
sources have a much better resolution ($\sim$0.8$^{\prime\prime}$), 
we can use this information and consider as 'good' pairs all those 
made of galaxies which are identified {in the Subaru/UDS images}, while we regard 
as potentially spurious those made of galaxies without such optical/near-IR 
counterparts. For the low-z sample, there are 11 pairs with distances 
between 0.001 and 0.003 degrees. Of these, only one was found to be closer 
than 5.4$^{\prime\prime}$ and both IRAC and R-band photometry indicate that we are 
dealing with two distinct sources. 10 close (i.e. again with distances 
between  0.001 and 0.003 degrees) pairs are instead found in the high-z sample. 
Of these, only one is at a distance below 5.4$^{\prime\prime}$. 
Unfortunately, the two sources of this pair are unidentified both in the 
optical and in the near-IR bands, and so the pair has to be considered as 
spurious.

The angular correlation function for the high-z sample was then computed by 
both including and excluding the possibly spurious pair. The small-scale 
($\theta \sim 0.0015 ^\circ$) results are shown by the two horizontal 
dashes in the right-hand panel of Figure 4. The filled circle represents 
their average. 
As it is possible to see, the $w(\theta)$ variation caused by the eventual 
presence of the spurious pair is almost negligible, and surely 
smaller than the error associated to the determination of $w(\theta)$. 
Nevertheless, in the following we will use as the 'best' small-scale point, that 
obtained as the average of the two estimates and the associated error will be 
the sum in quadrature of the Poisson one and of the variation due to the 
inclusion/exclusion of the pair.

If we then assume a power-law form for $w(\theta)=A\theta^{1-\gamma}$,
we can estimate the parameters $A$ and $\gamma$ by using a least-squares
fit to the data.  Given the large errors on $w$  
we choose to fix $\gamma$ to the standard value $\gamma=1.8$.
Although somewhat arbitrary, this figure and its assumed lack of dependence 
on redshift is justified by LSS observations of large enough samples of 
high redshift sources so to allow for a direct estimate of the slope of $w(\theta)$
at different look back times (e.g. Porciani, Magliocchetti \& Norberg 2004; 
Le Fevre et al. 2005).
The small area of the UDS survey introduces a negative bias through
the integral constraint, $\int w^{est} d\Omega = 0 $. We allow for
this by fitting to $ A \theta^{1-\gamma} -C $, where $C = 2.5 A$.\\
The dashed lines in Figure~4a and 4b represent the best power-law fits 
respectively to the low-z and high-z data. 
The associated best-fit values for the amplitude are $A^{\rm lz}
=[3.8\pm 1.6]\cdot 10^{-3}$ for the low-z sample, 
and $A^{\rm hz}=[10.0\pm 3.5]\cdot 10^{-3}$ for the 
high-z sample. 

The amplitude for the high-z sample is in good agreement 
with that obtained by Magliocchetti et al. (2007) by analyzing a sample of 
$\sim 800$ heavily obscured ($R\ge 25.5$) Spitzer-selected sources from the 
First Look Survey (FLS, Fadda et al. 2006) with 
$F_{24\mu \rm m}\ge 0.35$~mJy. This provides us with some confidence that 
cosmic variance is not a cause for main concern in our analysis of high-z 
UDS-SWIRE sources. Furthermore, as already noticed by the above 
authors, in this case $A$ is about four times higher than that derived by
Fang et al. (2004) for a sample of IRAC galaxies selected at
8$\mu$m ($A\sim 2.34 \cdot 10^{-3}$), and about ten times higher
than that obtained by Magliocchetti et al. (2007) for the 
whole $F_{24\mu \rm m} \ge 0.35$~mJy FLS dataset ($A=[9\pm 2]\cdot 10^{-4}$). 


\begin{table}
\begin{center}
\caption{Values for the angular correlation function $w(\theta)$ as obtained for the low-z 
and high-z samples. Error bars are obtained from Poisson statistics.}
\begin{tabular}{lll}
 $\theta$ [Deg] &Low-z sample & High-z sample\\
\hline
0.0015 & $1.9 \pm 0.9$ &    $3.8\pm  1.7$   \\
0.0045 & $0.6 \pm 0.4$  & $0.8 \pm 1.8$\\
0.0075 & $-0.19 \pm 0.25$  & $0.5 \pm  0.6$\\
0.0150 & $0.10  \pm 0.07$ & $0.2\pm 0.1$ \\
0.0450 & $0.07\pm 0.04$ & $0.05 \pm 0.08$\\
0.0750 & $-0.004 \pm 0.03$ & $0.08\pm 0.05$\\
0.1500 & $-0.0004 \pm 0.01$ & $0.03\pm 0.02$\\
0.2500 & $-0.01\pm 0.01$ & $-0.013\pm 0.017$
\end{tabular}
\end{center}
\end{table}

The case for the low-z sample is more tricky to deal with. 
The inferred value for the amplitude of its angular correlation function in 
fact suggests that UDS-SWIRE sources at $z\sim 1$ are sensibly less clustered 
(about a factor three in angular signal) than higher redshift ones. 
However, the computed $w(\theta)$ for the low-z sample presents puzzling negative values in four  
$\theta$ bins which should instead correspond to $w(\theta)> 0$. We have re-analyzed the sample 
a number of times by both varying the bin size and by adopting different prescriptions and dimensions for the 
random catalogue in the calculation of $w(\theta)$ and we can conclude that the 
observed negative values for $w(\theta)$ are indeed a real feature of the $z\sim 1$ sample. 
However, we cannot quantify how much of this trend is affected by poor statistics and/or 
cosmic variance (for instance, Gilli et al. 2007 find a variation in the number of $z\sim 1$, 24$\mu$m-selected sources brighter than approximately our chosen limit of almost a factor 2 between 
the two GOODS fields set in the northern and in the southern sky). While we note that 
also the Gilli et al. (2007) correlation function as estimated at $z\sim 0.7$ 
for the GOODS South exhibits a dip at scales $\sim 0.2/h$~Mpc which -- for the chosen cosmology -- 
correspond to the $\theta\sim 0.008^\circ$ bin where we find a negative value for $w(\theta)$, 
nevertheless caution has to be used when dealing with the four negative $w(\theta)$ figures. 
This is why we have decided to lower the weight of these points in our $\chi^2$ 
analysis by considering the estimated $w(\theta)+1\sigma$ values (see Table~1) as mere upper limits.
The resulting amplitude of the angular correlation function for low-z UDS-SWIRE galaxies 
in this case is $A^{\rm lz}=5.5^{+1.5}_{-2.0}\cdot 10^{-3}$, about 40 per cent higher than what previously 
found by including measured data points for $w(\theta)$ at all scales. This is the value which we will 
adopt throughout the paper and which -- though allowing for the large uncertainties -- still 
results to be a factor $\sim$2 lower than that derived from the high-z sample. We will investigate 
the implications of this result in the following sections.

\subsection{Relation to spatial quantities}
\begin{table*}
\begin{center}
\caption{Properties of the two $F_{400\mu\rm m}\ge 400$~mJy samples considered 
in this work. From left to right the columns show the average redshift 
$<z>$, the number of sources, the comoving number density $\bar{n}_G$ (in 
$[10^{-5}\:$Mpc$^{-3}]$ units), the amplitude $A$ of the angular correlation 
function, the comoving clustering length $r_0$ (in Mpc units), the minimum mass $M_{\rm min}^{\rm NFW}$ 
(in $M_\odot$ units) for the dark matter halos, normalization $N_0^{\rm NFW}$ and slope of the HON 
$\alpha^{\rm NFW}$ -- all three estimated under the assumption of a NFW profile for the distribution 
of galaxies within their dark matter halos --, the same quantities evaluated under the assumption of 
a steeper $\rho\propto r^{-3}$ (PL3) profile, and the large-scale bias $b$.}
\begin{tabular}{llllllllllll}
 $<z>$ & N & $\bar{n}_G$& $A \cdot 10^{-3}$ & $r_0$ & $\rm Log M_{\rm min}^{\rm NFW}$& $\rm Log N_0^{\rm NFW}$& $\alpha^{\rm NFW}$ & $\rm Log M_{\rm min}^{\rm PL3}$& $\rm Log N_0^{\rm PL3}$& $\alpha^{\rm PL3}$&bias\\
\hline
2.02  & 210 & $2.5\pm 0.3$ & $10.0\pm 3.5$& $15.9^{+2.9}_{-3.4}$& $12.8^{+0.3}_{-0.3}$&$-0.7^{+0.5}_{-0.5}$& $0.8^{+0.3}_{-0.6}$ &$12.8^{+0.4}_{-0.2}$& $-0.7^{+0.5}_{-0.5}$& $0.7^{+0.3}_{-0.5}$&6.17\\ 
0.79 & 350 & $9.9 \pm 1.0$ & $5.5^{+1.5}_{-2.0}$& $8.5^{+1.5}_{-1.8}$& $11.9^{+0.6}_{-0.8}$& $-2.0^{+0.8}_{-0.9}$& $0.8^{+0.1}_{-0.2}$&$11.8^{+0.8}_{-0.8}$&$-2.1^{+1.1}_{-0.9}$&$0.7^{+0.1}_{-0.3}$&1.70 \\ 
\end{tabular}
\end{center}
\end{table*}

The standard way of relating the angular two-point correlation
function $w(\theta)$ to the spatial two-point correlation function
$\xi(r,z)$ is by means of the relativistic Limber equation (Peebles,
1980):
\begin{eqnarray}
w(\theta)=2\:\frac{\int_0^{\infty}\int_0^{\infty}F^{-2}(x)x^4\Phi^2(x)
\xi(r,z)dx\:du}{\left[\int_0^{\infty}F^{-1}(x)x^2\Phi(x)dx\right]^2},
\label{eqn:limber} 
\end {eqnarray}
where $x$ is the comoving coordinate, $F(x)$ gives the correction for
curvature, and the selection function $\Phi(x)$ satisfies the relation
\begin{eqnarray}
{\cal N}=\int_0^{\infty}\Phi(x) F^{-1}(x)x^2 dx=\frac{1}{\Omega_s}
\int_0^{\infty
}N(z)dz,
\label{eqn:Ndense} 
\end{eqnarray}
in which $\cal N$ is the mean surface density on a surface of solid angle
$\Omega_s$ and $N(z)$ is the number of objects in the given survey
within the shell ($z,z+dz$). If we adopt a spatial correlation function 
of the form $\xi(r,z)=(r/r_0)^{-1.8}$, independent of redshift
 in the considered intervals, and we consider the 
redshift distributions presented in Figure~\ref{figure:nz}, 
for the adopted cosmology we obtain $r_0^{\rm hz}=15.9^{+2.9}_{-3.4}$~Mpc and 
$r_0^{\rm lz}=8.5^{+1.5}_{-1.8}$~Mpc (where both quantities are comoving and this latter value 
has been calculated for an amplitude $A=0.005$ found by following the method explained in \S3.1), respectively for the high-z and 
low-z sample. In order to test for the constancy of $r_0$ in the somehow wide 
redshift range $z=[1.6-3.5]$, we have also computed $w(\theta)$ and $r_0$ by using 
a subsample of 178 high-z galaxies within the narrower interval $z=[1.6-2.4]$.  
The results for both $A=0.011\pm 0.004$ and $r_0\sim 15$~Mpc are virtually identical 
to those reported here and in \S3.1.

Even though allowing for the large uncertainties, the above figures indicate that the high-z sample is more 
strongly clustered than the low-z one. In fact, 
despite the large error-bars, the inferred correlation lengths are incompatible 
with each other at the $\sim 3\sigma$ confidence level. The $r_0$ value for the high-z 
sample perfectly matches that of $\sim 15$~Mpc obtained by 
Magliocchetti et al. (2007). This is also in agreement with the 
estimates obtained in the case of ultra-luminous infrared galaxies over $1.5
\simlt z \simlt 3$ (Farrah et al. 2006a,b). Spitzer-selected galaxies found in the range 
$1.5\simlt z\simlt 3$ thus appear  to be amongst the most strongly clustered 
sources in the Universe. Similar values have been recently obtained by 
Foucaud et al. 
(2007) for their $z\sim [1-2]$ UKIDSS-UDS sample of distant red 
galaxies (DRG). Although with larger uncertainties, Grazian et al. 
(2006) also report a correlation length of the order of $13/h$~Mpc for their 
DRG, $2\simlt z\simlt 3$ dataset. A very high 
clustering length was also found by Magliocchetti \& Maddox (1999) in their 
statistical analysis of high redshift galaxies in the Hubble Deep Field 
North. All these values point to an evolutionary connection 
between galaxies undergoing an intense star-formation activity such as those probed by 
Spitzer and the HDF beyond $z\sim 2$ and older sources such as distant red 
galaxies. According to this picture, these two classes of objects 
would merely correspond to different stages in the evolution of a very massive 
galaxy. We note that locally, the clustering properties of $z\sim 2$ active galaxies find a
counterpart only in those exhibited by radio sources (see e.g.
Magliocchetti et al. 2004) and early type $L\simgt 4\;L^*$ 
galaxies (see e.g. Norberg et al. 2002) and are only second to those of rich
clusters of galaxies (e.g. Guzzo et al. 2000).  It is then natural trying to 
envisage a connection between these latter objects with the high-z ones, 
whereby massive galaxies undergoing intense star-formation at redshifts 
$z\sim 2$ end up as the very bright central galaxies 
(passive objects with a high 
probability for enhanced radio activity, see e.g. Best et al. 2007; 
Magliocchetti \& Bruggen 2007) of local clusters.

On the other hand, Spitzer-selected sources residing at $z\sim 1$ are
less strongly clustered. Interestingly enough, both Magliocchetti \& Maddox (1999) and 
Grazian et al. (2006) find similar results for their 
$1\simlt z \simlt 2$ HDF and DRG samples. Our results for the low-z sample also closely 
resemble those recently obtained by Gilli et al. (2007) 
who find a correlation length $r_0\sim 8$~Mpc for their joint sample of 
$0.5\simlt z\simlt 1$, $L_{\rm IR}> 10^{11}L_\odot$  galaxies selected at 24$\mu$m 
in the two GOODS fields.\\
Locally, the clustering properties of mid-IR bright sources 
mirror those of 'normal' early-type galaxies (e.g. Madgwick et al. 2003; Zehavi et al. 2005).    

\section{Connection with Physical Properties}
A closer look at Figures 4a and 4b shows that a simple
power-law cannot provide a good fit to the measured $w(\theta)$ 
below $\theta\simlt 0.003^\circ$. 
This small-scale steepening is intimately related to the way the
sources under consideration occupy their dark matter haloes, an issue which
can be dealt with within the so-called Halo Occupation Scenario. 
For details, we refer the interested reader to the work of Porciani, 
Magliocchetti \& Norberg (2004). Basically, the halo occupation 
framework relates the clustering properties of a chosen population 
of extragalactic objects with the way such objects populate their dark matter 
haloes. Within this scenario, the two-point correlation function $\xi$ can be 
written as the sum of two components, $\xi^{1h}$ and $\xi^{2h}$, where the 
first quantity accounts for pairs of galaxies residing within the same halo, 
while the second one takes into account the contribution to the correlation 
function of galaxies belonging to different haloes. 
The quantities $\xi^{1h}$ and $\xi^{2h}$ depend on a number of factors, 
amongst which cosmology, spatial distribution of sources within their haloes, 
mass function of the dark matter halos, dark matter auto-correlation function, 
large-scale bias, and also on the 
first $\langle N\rangle$ and second $\sigma$ moments of the halo occupation 
distribution $P_N(M)$ which gives the probability of finding $N$ galaxies
within a single halo as a function of the halo mass $M$. On the other hand, 
for a given cosmology, given the halo mass function $n(M)$ (number density of dark matter haloes
per unit comoving volume and $\log_{10}(M)$), the first moment of
the halo occupation distribution $\langle N\rangle$ (hereafter called halo 
occupation number or HON) completely determines the 
mean predicted comoving number density $\bar{n}$ of galaxies in the desired 
redshift range:
\begin{equation}
\bar{n}=\int n(M)\,\langle N(M)\rangle \,dM\;. \label{eq:avern}
\end{equation}
Any consistent picture within the halo occupation scenario has to be able to 
simultaneously reproduce at least the first (mean number density) and second 
(two-point correlation function) moment of the observed galaxy distribution.

One common way to parametrize the HON and variance of the galaxy 
distribution is (Porciani et al. 2004; see also Hatton et al. 2003):
\begin{eqnarray}
N(M) = \left\{\begin{array}{ll} N_0(M/M_{\rm min})^{\alpha} &
\rm{if}\ M \ge  M_{\rm min} \\
0 & \rm {if} \ M<M_{\rm min}
\end{array};\right.\nonumber\\
\label{eq:Ngal}
\sigma(M)=\beta(M)^2 N(M),\;\;\;\;\;\;\;\;\;\;\;\;\;\;\;\;\;\;\;\;\;\;\;\;\;\;\;\;\;\;\;
\end{eqnarray}
where $\beta(M)=0,{\rm log}(M/M_{\rm min})/{\rm log}(M/M_0),1$,
respectively for $N(M)=0$, $N(M)<1$ and $N(M)\ge 1$. The
operational definition of $M_0$ is such that $N(M_0)=1$, 
while $M_{\rm min}$ is the minimum mass of
a halo able to host a source of the kind under consideration. More
and more massive haloes are expected to host more and more
galaxies, justifying the assumption of a power-law shape for the
halo occupation number. As for the variance $\sigma(M)$, we note that 
the high-mass value for $\beta(M)$ simply reflects the Poisson statistics, 
while the functional form at intermediate masses (chosen to fit the results
from semi-analytical models and hydrodynamical simulations, 
see e.g. Sheth \& Diaferio 2001 and Berlind et al. 2003) 
describes the (strongly) sub-Poissonian regime.

\begin{figure*}
\includegraphics[width=8.0cm]{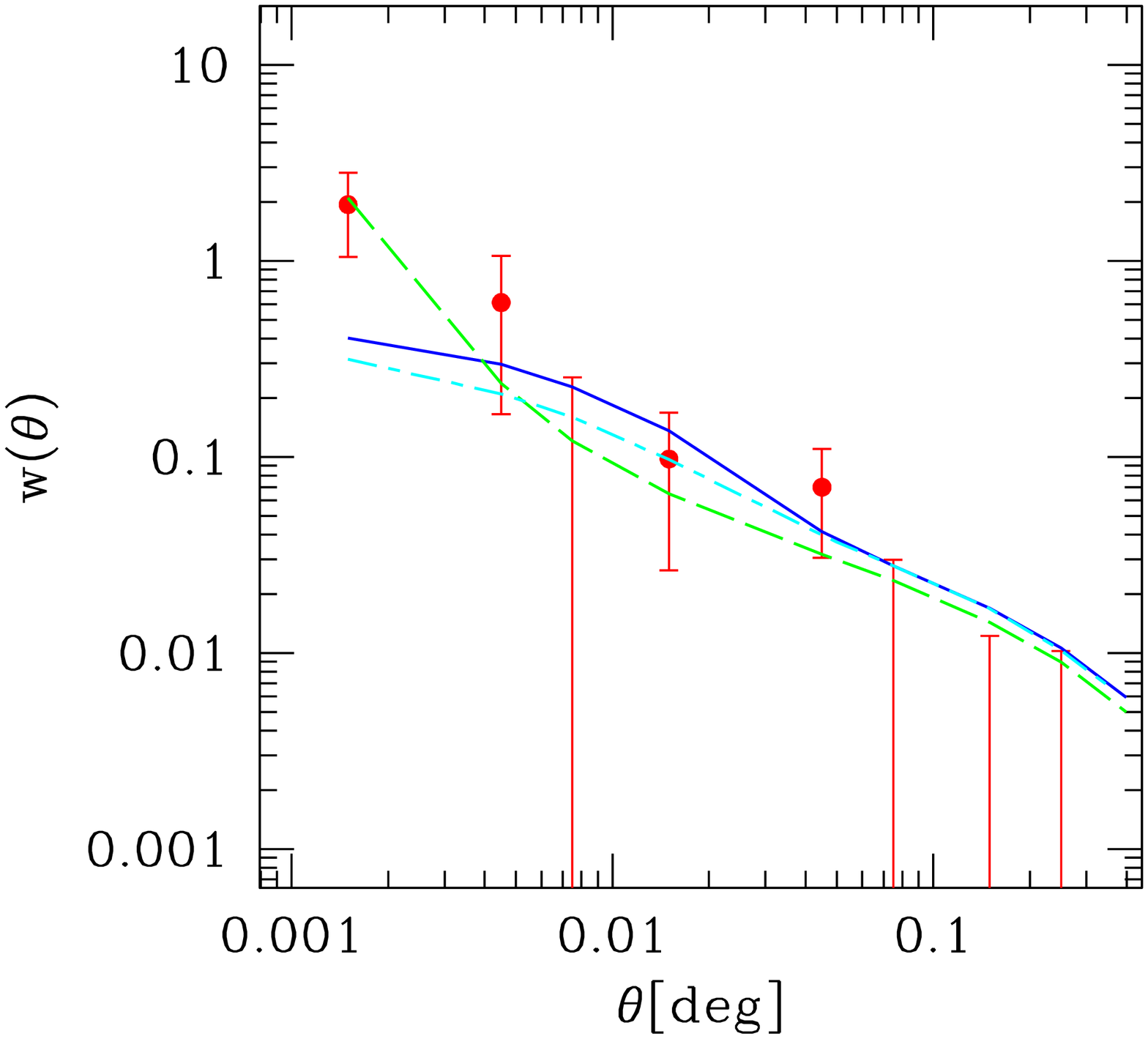} 
\includegraphics[width=8.0cm]{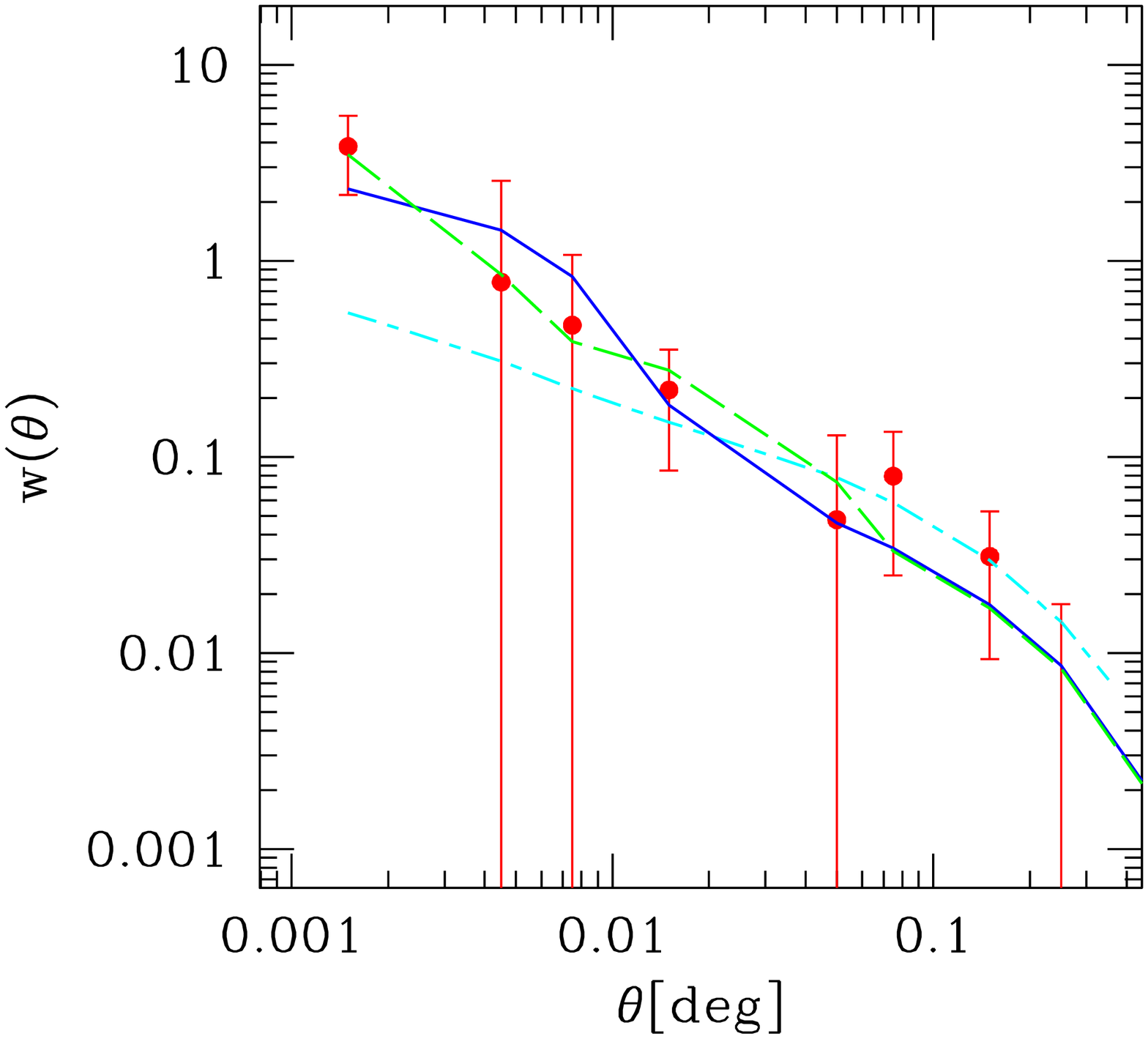}
\label{figure:wmod}
\caption{Angular correlation function $w(\theta)$ for the low-z 
(left-hand panel), and high-z (right-hand panel) samples. The solid (blue) curves represent the best 
HON fits to the data obtained under the assumption that the distribution of galaxies 
within their dark matter haloes mirrors that of the dark matter (chosen to be described by a 
NFW profile, cfr. Section 4), while the dashed (green) curves are the best fits obtained for a steeper 
galaxy profile, $\rho \propto r^{-3}$. The long-short dashed (cyan) curves correspond to the best 
1-parameter fits obtained by setting in eq. (5) $\alpha=0$ and $N_0=1$. 
The negative $w(\theta)$ data-points in the $\chi^2$ analysis of the low-z sample have been treated as in \S3.1 
(see text for details).}
\end{figure*}

In more operative terms, we have calculated $\xi^{1h}$ and $\xi^{2h}$ 
using the Sheth \& Tormen (1999) prescriptions for the halo mass function 
and the large-scale bias, while the mass auto-correlation function was computed by 
following a revised version of the method of Peacock \& Dodds (1996) which takes into 
account the spatial exclusion between haloes (i.e. two haloes cannot occupy the same 
volume). As a starting working hypothesis, the radial profile of the galaxy distribution 
within their halos is assumed to follow that of the dark matter for which 
we adopt a NFW profile (Navarro, Frenk \& White, 1997). Finally, 
we allowed the parameters in eq.~(\ref{eq:Ngal}) to vary within the
following ranges:\\
 $0 \le \alpha\le 2$; $10^{11}\le M_{{\rm min}}/M_\odot \le 10^{14}$; 
$-3\le \log_{10}(N_0)\le 0$.

Values for these three parameters have been then determined through a
minimum $\chi^2$ technique by simultaneously fitting the 
observed $w(\theta)$ (where for the low-z sample we have treated the negative $w(\theta)$ values as in \S3.1) 
and the estimated number density $\bar{n}_G$ of sources in both 
the low-z and high-z samples (cfr. Table~2).
The angular correlation function was computed from
eq.~(\ref{eqn:limber}), with $\xi$ obtained as explained above and the 
redshift distributions shown in Figure ~1. 
The best-fit parameters for the HON of both the low-z and high-z sample 
are summarized in Table~2; the quoted errors correspond to $\Delta\chi^2=1$.
The uncertainties associated with $\alpha$ and especially to $M_{\rm min}$ are much smaller than 
those corresponding to the normalization parameter $N_0$ as while the first two quantities are 
constrained by both the large-scale and small-scale behaviour of $\xi$, according to our model 
$N_0$ only enters the description of the spatial two-point correlation function in the sub-halo regime 
(cfr. eqs 13 and 14 of Porciani et al. 2004).

The theoretical angular correlation functions corresponding to the
best-fit HON parameters for both the low-z and high-z sample are represented by the 
solid (blue) curves in Figure~5. Since the high-z sample includes a fraction of sources without 
estimated redshifts, we have also re-run the HON method by only including 
in the calculation of the comoving number density $\bar{n}_G$ sources with assigned photo-z. 
The results are virtually identical to those reported in Table~2.

As it is possible to notice from Figure~5, the HON does a good job at reproducing the observed 
$w(\theta)$ at intermediate-to-large angular scales both in the case of the low-z sample and for 
the high-z sample. However, as it is particularly evident 
in the low-z plot, it cannot successfully describe the steep rise of the measured two-point 
correlation function below $\sim 0.005$ degrees. The reason is quite easy to understand: 
strong contributions to the 1-halo correlation function $\xi^{1h}$ (i.e. the portion of $\xi$ 
which determines its behaviour on scales smaller than the virial radius) are mainly due 
to high values of $\alpha$ and $N_0$. The magnitude of $\alpha$, together with that of 
$M_{\rm min}$, is however also determined by the large-scale behaviour of $w(\theta)$. 
Furthermore, all three quantities enter the calculation of the predicted number density 
$\bar{n}$ via equation (\ref{eq:avern}) and the number of possible values and combinations is 
strictly ruled by the requirement that the predicted $\bar{n}$ matches (within 2$\sigma$ 
in our case) the observed, $\bar{n}_G$, one.  It follows that the best-fit curves plotted in 
Figures 5a and 5b below $\theta \sim 0.03^\circ$ in the low-z case and 
$\theta\sim 0.01^\circ$ for the high-z sample represent the largest possible small-scale 
contributions to the observed 
$w(\theta)$ subject to the constraints put by the large-scale behaviour of $w(\theta)$ 
and by the observed number density of the sources that produce the clustering signal. 
Furthermore, we note that large values for $\alpha$ and $N_0$, as well as more extreme 
choices for the second moment $\sigma(M)$ of the galaxy distribution in (5), 
would only determine a boost
of $w(\theta$) on the smallest angular scales probed by our analysis (i.e. an approximately 
vertical shift of the projected contribution of $\xi^{1h}$), but cannot radically change 
its shape, which shows a smooth rise followed by a plateau in net contrast with the steep 
jump of the observed data points on scales smaller than at least $\theta\sim 0.003$ degrees.
The model as it is cannot do any better than this.

The most likely solution to this 'small-scale crisis' comes from releasing the 
assumption that the distribution of galaxies within their halos follows that 
of the dark matter. This working hypothesis was used both for simplicity and also because 
locally the halo distribution of both late-type and early-type galaxies can be nicely described by 
a NFW profile (see e.g. Magliocchetti \& Porciani 2003). However, this does not have to be 
necessarily true at high redshifts. As an alternative approach, we have then considered the  
possibility for the galaxies to be more concentrated than the underlying dark matter, and in 
order to reproduce this effect we have taken a galaxy density profile of the kind 
$\rho\propto r^{-3}$ (hereafter called PL3; see Magliocchetti \& Porciani 2003).

The best fitting curves obtained within the HON framework in the PL3 case are shown in Figures 5a 
and 5b by the dashed (green) curves. The corresponding best-fitting values for $M_{\rm min}$, 
$\alpha$ and $N_0$ are given in Table~2. As expected, a model which now foresees a steep profile 
for the galaxy distribution can provide an excellent fit to the observed small-scale rise of 
$w(\theta)$ in both the low-z and high-z sample.  Furthermore, the best fitting values for the 
HON in the PL3 case are almost identical to those obtained by considering NFW profiles. This shows 
that the average behaviour for the spatial distribution of galaxies within their dark matter 
halos can be recovered from investigations of the small-scale features of the angular/spatial 
two-point correlation function in a 'clean' way as such features can be modelled  
independently of the other parameters adopted to describe the 
clustering properties of a given astrophysical population. 

So, which kind of information can we infer from this small-scale behaviour of $w(\theta)$? 
Locally, Magliocchetti \& Porciani (2003) find that galaxy distributions of the PL3 kind 
are strongly rejected by the data, as small-scale measurements of the two-point correlation 
function prefer smoother profiles such as NFW or $\rho \propto r^{-2.5}$. However, at very low 
redshifts it is reasonable to expect the majority of dark matter halos to be virialized 
and therefore be to reasonably described (at least beyond the very small scales, see e.g. Bukert 1995) 
by NFW-like profiles.
Two options are instead possible at such high redshifts. The first one is that galaxies 
still do trace the underlying distribution of the dark matter within their halos, and it is the 
dark matter itself which does not follow NFW-like profiles. The second option is 
instead that galaxies are simply more radially concentrated than the dominant dark matter 
counterpart. Whatever the favourite choice, one thing is clear. On sub-halo scales, 
24$\mu$m-selected galaxies at redshifts $z\simgt 0.6$ are more strongly radially concentrated than 
their low-redshift counterparts. Such a strong concentration is indicative of a strong 
interaction between galaxies which reside in the proximity of the halo centres, and 
we expect such a strong interaction to eventually drive a substantial number of mergers. 
It is possibly not a random coincidence that bright 24$\mu$m-selected sources at 
intermediate-to-high redshifts are in general associated to very intense star-forming 
systems; the excess of proximity between such galaxies would eventually 
lead to gas-rich mergers which could easily trigger enhanced star-formation activity like 
the ones observed. In passing, we note that evidence for an upturn (although less 
significative than ours) on scales $r\simlt 0.2/h$~Mpc has also been recently reported by Gilli et al. (2007) 
for their sample of $z\sim 1$, 24$\mu$m-selected sources in the GOODS fields. 

\begin{figure}
\includegraphics[width=8.0cm]{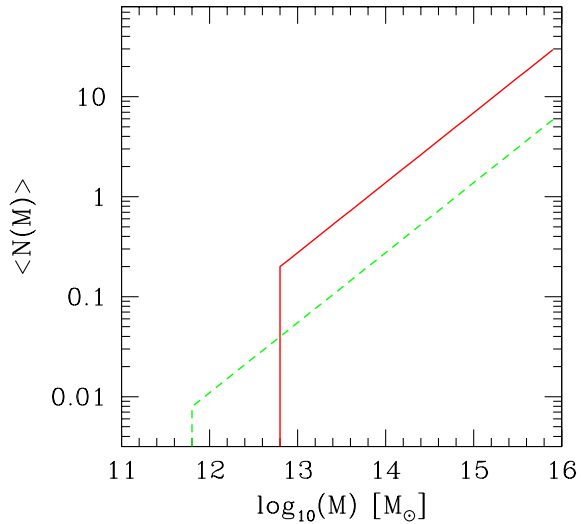} 
\caption{Average number of galaxies $\langle N\rangle$ per dark matter halo 
of specified mass 
$M$ (expressed in $M_\odot$ units). The solid (red) line represents the case for 
the high-z sample, while the dashed (green) line is for the low-z sample.}
\end{figure}

\begin{figure}
\includegraphics[width=8.0cm]{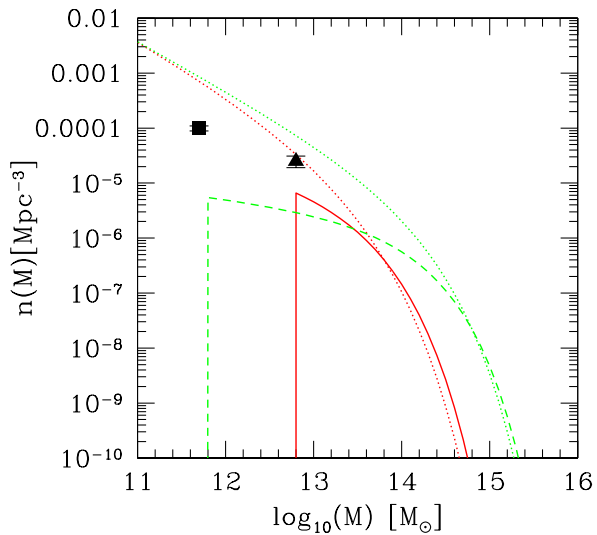}
\caption{Number of galaxies per unit of (log) dark matter mass and volume as evaluated following 
eq.(4) for the best fitting values of the HON presented in Figure 6 at the effective 
redshifts $z\sim 0.8$ and $z\sim 2$ of the two datasets under exam. 
The solid (red) line represents the result for 
the high-z sample, while the dashed (green) line is for low-z sample. 
For comparison, the two dotted lines indicate the Sheth \& Tormen (1999) mass 
function $n(M)$ of dark matter haloes (the higher/green curve corresponding to the low-z sample). 
The filled triangle indicates the observed 
number density of sources in the high-z sample, while the filled square is for the low-z sample.}
\end{figure}

The best average number of galaxies $\langle N\rangle$ as a function of halo mass for the two 
analysed samples (and in the PL3 case) are presented in  Figure~6 (solid/red 
line for the high-z sample), while the galaxy mass function as coming from 
eq. (4) for the best-fit values of the HON for the high-z sample (solid/red curve) 
and in the case of the low-z sample (dashed/green curve) are illustrated by Figure~7. 
Despite the large uncertainties due to the relatively poor statistics, 
a number of interesting features can be gathered from these plots. 
Sources belonging to the high-z sample appear to be exclusively located in  very massive, 
$M\simgt 10^{13}M_\odot$, structures, identifiable with groups-to-clusters of galaxies. 
These galaxies are quite common within their halos ($\langle N \rangle$ between $\sim 0.6$ and $\sim 
0.06$ at the lowest possible masses where the sources start appearing), and we note that since 
$\langle N \rangle$ is a quantity which is 
averaged all over the dark matter halos more massive than $M_{\rm min}$, in the most extreme case 
our results imply that more than {\it one in two} of all the 
structures with masses $M\sim 10^{13} M_\odot$ host a high-z, $F_{24 \mu \rm m}\ge 400 \mu$Jy galaxy.  
Furthermore, their number sensibly increases 
with the richness of the structure which hosts them. At around $M\sim 10^{15} M_\odot$, their average 
abundance is between 5 and 20 sources per halo ($\sim 10$ for the best-fit model), value which is not in  
disagreement with the typical figures for the number of bright, $L> L^*$ galaxies associated to local 
rich clusters (e.g. Terlevich, Caldwell \& Bower, 2001). 

On the other hand, the low-z sample seems to be made of galaxies of much smaller mass 
($M_{\rm min}\simeq 10^{12} 
M_\odot$). Also, these objects are very rare (in the most unfavourable case we get 
$\langle N \rangle\sim 10^{-3}$ 
at the lowest possible masses at which the sources start appearing, i.e. only 0.1\% of the 'allowed' 
halos host a galaxy of the kind which make the low-z sample), even though their number -- 
as it was the case for the high-z sample -- 
sensibly increases with halo mass; this determines an overall flatness of the galaxy mass 
function at least up to $\sim 10^{14} M_\odot$.

\begin{figure}
\includegraphics[width=8.0cm]{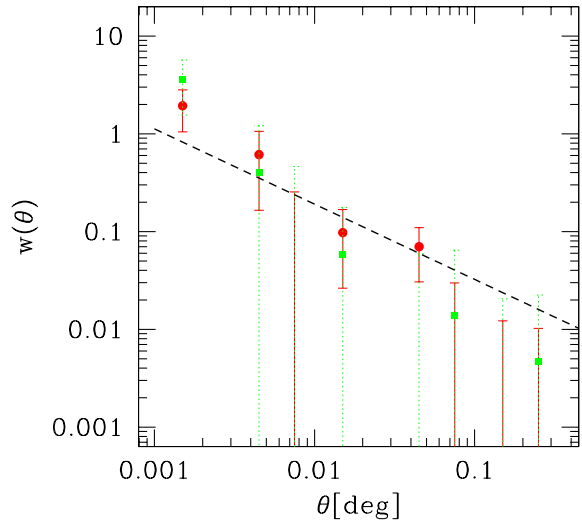}
\caption{Angular two-point correlation function $w(\theta)$ for low-z UDS-SWIRE sources
brighter than 0.5~mJy (210 objects; green squares and associated, dotted, errorbars) as
compared to that presented in \S3.1 (red circles). The dashed line represents the
best-fit to the low-z, $F_{24\mu\rm m}\ge 0.4$~mJy, sample obtained as in \S3.1}
\end{figure}

\noindent
From the analyses performed in Sections 3 and 4 it appears clear that the populations 
probed by the high-z and low-z samples are very different
from each other. One might argue that the higher clustering level reported at 
redshifts $z\simgt 1.6$ can be simply attributed to the fact that since we are dealing 
with a flux-limited  survey rather than a volume-limited one, higher redshifts probe 
more intrinsically luminous 
sources which are in general hosted by more massive halos and therefore result more strongly clustered.  
However, we note that this cannot be the only explanation for such a strong difference between 
the clustering properties of the low-z and the high-z sample. Sources in the high-z sample are in fact 
so strongly clustered that it would only take a small fraction of them 
in the low-z sample (say of the order of 20\%) to sensibly boost the low-z clustering signal. This does not 
seem to happen. Obviously, the ultimate evidence for a strong evolution of the clustering signal of 
24$\mu$m-selected sources between $z\sim 2$ and $z\sim 1$ could in principle only be obtained by comparing 
volume-limited rather than flux-limited samples.
Given the Spectral Energy Distributions of these sources, this would imply including in our analysis 
 only objects brighter than $\sim 1$~mJy in the low-z sample. Unfortunately, the number of such low-z 
sources is very limited and insufficient to allow for a statistically significant estimate of 
$w(\theta)$. However, as a sensible alternative we can consider a sub-sample of low-z sources 
brighter than the chosen 0.4~mJy limit, which contains enough objects to provide meaningful results 
for the two-point correlation function. We have then decided to restrict our analysis to low-z galaxies 
brighter than $F_{24\mu\rm m}=0.5$~mJy. We note that, despite the marginal flux increment, this new sample 
only contains 210 sources, i.e. almost a factor 2 less than the original, $F_{24\mu\rm m}\ge 0.4$~mJy, 
dataset. The corresponding $w(\theta)$ is shown by the (green) squares in Figure 8. \\
Under the assumption of no cosmological evolution for those 24$\mu$m-selected galaxies 
residing at $z\sim 2$ and taking into account the different volumes occupied by the two low-z and high-z 
samples, we would expect $\sim 80$ very massive sources in the $0.6\le z\le 1.2$ redshift interval. 
And if these sources indeed existed at $z\sim 1$, they would constitute $\sim 40$\% of the 
$F_{24\mu\rm m}\ge 0.5$~mJy low-z sample, to be compared to $\sim 25$\% of the $F_{24\mu\rm m}\ge 0.4$~mJy 
one. This sensible increase in the relative weight of very massive sources (which now 
would make only slightly less than 50\% of the entire low-z dataset) in the $z\sim 1$ sample  
is expected to determine a sensible boost of the correlation function signal.
But this is does not happen, as the observed clustering amplitude of $F_{24\mu\rm m}\ge 0.5$~mJy objects 
is virtually identical to that obtained in \S3.1 (and reproduced in Figure 8 by the red 
circles for sake of clarity) for a lower flux cut.

Massive galaxies undergoing intense activity such as star-formation or AGN accretion seem 
to disappear when going from redshifts $\sim 2$ to 
$\sim 1$. At this lower redshifts, obscured star forming and AGN activity seem to be segregated in much 
lower-mass systems. Obviously, the present data cannot say anything on the eventual presence of active 
low-mass galaxies at redshift $\sim 2$, but they tend to strongly exclude the presence of intense activity 
from very massive systems below redshifts - say - 1.5. 

\subsection{The Halo Bias Approach}
One might wonder about the need of using somewhat sophisticated tools 
such as the Halo Occupation formalism in the case of datasets characterized by a low 
signal-to-noise ratio as the ones we are dealing with. To answer this question, 
we have considered the more standard {\it halo bias} approach (Mo \& White 1996) 
which describes the clustering signal of a chosen population of extragalactic sources 
as the product between the two-point correlation function of the dark matter and the square of the so-called 
bias function, a quantity which solely depends on the minimum mass $M_{\rm min}$ 
of the halos in which the astrophysical objects reside. Such an approach assumes a one-to-one 
correspondence between dark matter halos and hosted sources and can be thought as a special case 
within the HON framework corresponding in equation (5) to $\alpha=0$, $N_0=1$ and $\sigma(M)=0$. 
Once the cosmology is fixed, this is a 1-parameter model which might seem more appropriate 
for the description of clustering signals affected by large noise. 
The angular two-point correlation functions $w(\theta)$ predicted by the halo bias model 
 for the low-z and high-z samples have then been computed once again by following eq.(2) and the 
best values for $M_{\rm min}$ in the two cases have been found by a $\chi^2$-fit to the 
observed clustering signal. We find ${\rm Log}M_{\rm min}=13.3^{+0.2}_{-0.4}$ and 
${\rm Log}M_{\rm min}=12.4^{+0.4}_{-0.6}$ respectively for the high-z and low-z sample, 
and the corresponding $w(\theta)$'s are represented in Figures 5a and 5b by the (cyan) long-short dashed curves.
These values are in good agreement with those obtained by using the full HON approach presented 
earlier in this Section, the somewhat higher figures found for $M_{\rm min}$ in this latter case being 
simply due to the fact that $\alpha$ and $M_{\rm min}$ are covariant quantities, so that within the 
HON framework higher $\alpha$ in general correspond to lower $M_{\rm min}$. This provides reassuring 
evidence for the goodness of the results derived from the HON analysis. 

On the other hand, the 1-parameter 
halo bias model suffers from a number of problems which can instead be overcome by using the full HON 
approach. The first problem is encountered once the values of $M_{\rm min}$ as obtained above are plugged 
in the calculation of the number density of sources via eq. (4). In the high-z case, the figure  
we obtain ($\bar{n}\sim 1.5 \cdot 10^{-5}$~Mpc$^{-3}$) is in good agreement with the observational findings 
presented in \S2. However, for the low-z sample we find $\bar{n}\sim 8.5 \cdot 10^{-3}$~Mpc$^{-3}$, 
value which is about 10 times larger that what observationally determined. 
This big discrepancy stems from the erroneous {\it a priori} assumption of the 1-parameter halo bias model 
for a one-to-one correspondence between dark matter halos and astrophysical sources. As shown by the results 
of the full HON analysis (cfr. Table~2), such a working hypothesis could hold for the high-z sample which 
presents $N_0$ values not too far from unity, 
but badly fails when one deals with rare sources 
such a those characterizing the low-z sample. The HON approach instead provides a self-consistent 
answer for both the clustering properties and the number density of $z\sim 1$ sources.  

A second problem related to 1-parameter models such as the halo bias 
is that, by merely concentrating on the clustering properties of the dark matter halos, they
are unable to describe the clustering behaviour of astrophysical sources on sub-halo scales.
This feature is evident in Figures 5a and 5b. Although reproducing the large-scale trend of the observed 
$w(\theta)$'s on large scales, the halo bias model systematically underestimates the observed 
clustering signal at small angular distances. The problem is particularly bad at the smallest 
angular scales probed by our analysis whereby, in the case of $z\sim 2$ sources, this model 
predicts $\sim 3$ pairs to be compared to the 9 (or 10, see \S2) observed in our dataset. 
For the lower redshift sample the predicted figure is again $\sim 3$, while the number of observed pairs 
is 11. The full HON approach instead works much better at describing the clustering properties of 
extragalactic sources at all scales and can provide a correct match between observed and predicted 
quantities even at the shortest distances (the number of predicted pairs at $\theta\sim 0.0015$~degrees 
within the HON framework is $\sim 5$ and $\sim 9$ respectively for a NFW and a PL3 model at $z\sim 2$ 
and $\sim 4$ and $\sim 11$ at $z\sim 1$). We note, for comparison, that the '{$\gamma$} model' exploited in \S3, 
gives $N_{\it low-z}\sim 6$ and $N_{\it high-z}\sim 5$.

\section{Conclusions}
We have investigated the clustering properties of 24$\mu$m-selected galaxies brighter 
than 400$\mu$Jy, drawn from the SWIRE and UKIDSS Ultra Deep Surveys. With the help 
of photometric redshift determinations, we have concentrated on two datasets which include 
galaxies respectively in the $z=[0.6-1.2]$ ({\it low-z} sample) and $z\simgt 1.6$ ({\it high-z} sample) 
redshift ranges. The low-z sample is made of 350 galaxies, while the high-z sample includes both 182 
galaxies with estimated photo-z's and another 28 sources which do not have any optical or near-IR 
counterpart. Diagnostics based on the ratio between 8$\mu$m and 24$\mu$m fluxes indicate that 
these two samples are likely made by a very similar mixture of active star-forming galaxies 
($\sim 65$ per cent) and AGN (the remaining $\sim 35$ per cent). 

Results obtained by fitting with a power-law of fixed slope $\gamma=1.8$ the observed angular two-point 
correlation function $w(\theta)$ report an amplitude $A^{hz}=0.010\pm 0.0035$ and $A^{lz}=0.0055^{+0.0015}_{-0.0020}$ 
($w(\theta)=A\;\theta^{1-\gamma}$) respectively for the high-z and low-z samples. In more physical 
coordinates, the above results imply (comoving) correlation lengths for the spatial two-point correlation 
function $\xi$ (modelled as $\xi=(r/r_0)^{-\gamma}$) $r_0^{hz}=15.9^{+2.9}_{-3.4}$~Mpc and 
$r_0^{lz}=8.5^{+1.5}_{-1.8}$~Mpc, showing that the galaxies in the high-z sample are more strongly 
clustered that those found at lower redshifts.

A deeper insight on the above findings is provided by the so-called Halo Occupation 
Scenario which connects the clustering properties of a chosen population of astrophysical objects 
with some of their physical properties. The key quantity for this kind of analysis is 
the Halo Occupation Number (HON), i.e. the average number of galaxies enclosed in a halo of given mass 
$M$, which can be parametrized as $\langle N\rangle=N_0\left(M/M_{\rm min}\right)^\alpha$, 
with $M_{\rm min}$ minimum mass 
for a halo to host one of such galaxies and $N_0$ normalization factor which provides information 
on how common the galaxies under examination are.

At first we have adopted as a working hypothesis that Spitzer-selected galaxies were spatially 
distributed within their dark matter halos according to the dark matter (NFW) distribution. 
However, we find that such a model fails at reproducing the high amplitude of the observed $w(\theta$) signal 
on angular scales $\simlt 0.003$ degrees both in the low-z and high-z sample. A much better agreement 
with the data is obtained when we take the galaxies to be very concentrated towards their halo centres, e.g. 
by assuming a distribution of the kind $\rho\propto r^{-3}$.
Such a high concentration -- not observed in local data (Magliocchetti \& Porciani 2003) 
-- is suggestive of a strong interaction and close encounters between galaxies which reside in the 
proximity of the halo centres, and we expect such a strong interaction to eventually drive a 
substantial number of gas-rich mergers which could easily trigger the observed enhanced star-formation activity. 

Investigations of the best-fit values for the HON report 
values ${\rm Log}\left(M_{\rm min}^{hz}/M_\odot\right)=12.8^{+0.2}_{-0.4}$; ${\rm Log} N_0^{hz}=-0.7^{+0.5}_{-0.5}$; 
$\alpha^{hz}=0.7^{+0.3}_{-0.5}$ for the high-z sample and ${\rm Log}\left(M_{\rm min}^{lz}/M_\odot\right)=11.8
^{+0.8}_{-0.8}$; ${\rm Log} N_0^{lz}=-2.1^{+1.1}_{-0.9}$; $\alpha^{lz}=0.7^{+0.1}_{-0.3}$ in the case of 
low-z sources. The above figures indicate that galaxies belonging to the high-z sample are exclusively 
associated with very massive, $M\simgt 10^{12.8}M_\odot$, structures, identifiable with 
groups-to-clusters of galaxies. Furthermore, these galaxies are quite common within their halos 
(in the most extreme case, our results imply that more than {\it one in two} of all the 
structures with masses $M\simeq M_{\rm min}$ host a $z\sim 2$, $F_{24 \mu \rm m}\ge 400 \mu$Jy galaxy), and
their number sensibly increases with the richness of the structure which hosts them. 
On the other hand, the low-z sample seems to be made of galaxies of much smaller mass 
($M_{\rm min}\simeq 10^{11.8} M_\odot$). These objects are also very rare; 
in the most unfavourable case we get that only 0.1\% of the 'allowed' 
halos host a galaxy of the kind which make the low-z sample. 

Such a remarkable difference in the clustering and environmental properties of active sources 
as seen at $z\sim 2$ and $z\sim 1$  can be hardly attributed to the fact that 
high redshift galaxies are intrinsically brighter than their lower redshift counterparts (cfr. \S4). 
Indeed, despite the large uncertainties determined by the poor statistics of the considered datasets, 
our results indicate that the populations probed by the high-z and low-z samples are very different 
from each other: massive active galaxies seem 
to disappear when going from $z\sim 2$ to $z\sim 1$, and at this lower redshifts all the AGN 
and starforming activity appears to be segregated in much lower-mass systems (see also the results 
of Gilli et al. 2007 who find for their bright, $z\sim 0.7$, 24$\mu$m-selected 
galaxies in the GOODS fields a correlation length $r_0\sim 8$~Mpc.).
We stress that, while investigations of the luminosity function of 8$\mu$m-rest frame-selected 
sources have already established a strong luminosity evolution of sources between redshifts $\sim 2$ 
and $\sim 1$ (e.g. Caputi et al. 2007), clustering measurements provide a unique tool to determine 
the physical nature of such an evolution. Our work can in fact show that the strong differential 
evolution of the 24$\mu$m LF is not simply due to the fact that the less luminous sources at $z\sim 1$ 
are dimmed versions of the galaxies at higher-z (i.e. pseudo-passive evolution), but indeed has to be 
attributed to different populations of objects inhabiting different dark matter haloes and structures. 

The general picture which then emerges from the results of this paper points to a differential evolution for 
high-mass and low-mass systems. Massive systems form early in time at the rare peaks of the density field. 
Their intense activity, both in terms of star-formation and AGN accretion, is strongly favoured (if not triggered) 
by numerous close encounters and gas-rich mergers happening in the proximity of the centres of the 
massive/cluster-like structures in which these sources reside. Such an active phase is relatively short-lasted 
and already below $z\sim 1.5$ 
these objects evolve as optically passive galaxies, in most of the cases ending up as the supermassive galaxies 
which locally reside in cluster centres and which show a remarkable tendency for enhanced radio activity (see e.g. 
Best et al. 2006; Magliocchetti \& Br\"uggen 2007).
Lower ($M\simlt 10^{12}M_\odot$) mass systems are instead characterized by an active phase which lasts down to lower 
redshifts. Also in this case, the AGN and star forming activities seem to be favoured by the strong concentration 
of these sources towards their halo centres. These objects will eventually end up as 'typical' early-type galaxies 
which favour relatively (but not extremely) dense environments.

At present, studies on the evolutionary properties of dust-enshrouded active systems like those we have investigated 
in this work are hampered by the lack of surveys which can probe their peak of emissivity 
to low enough fluxes and also by the extreme difficulty of getting spectroscopic redshifts for such optically faint 
sources. 
The forthcoming advent of instruments such as Herschel, SCUBA-2 and ALMA will finally fill in this gap and 
provide the community with 'the ultimate truth' on most of the issues connected with galaxy formation and evolution.

\section*{ACKNOWLEDGMENTS}
Thanks are due to the referee for constructive comments, that
helped improving the paper. CS acknowledges STFC for financial support.

\end{document}